# A square cross-section FOV rotational CL (SC-CL) and its analytical reconstruction method

Xiang Zou[a,b], Wuliang Shi[a,b], Muge Du[a,b], Yuxiang Xing[a,b]

[a]Department of Engineering Physics, Tsinghua University, Beijing, 100084, China;

[b]Key Laboratory of Particle & Radiation Imaging (Tsinghua University), Ministry of Education, Beijing, 100084, China

## ABSTRACT

Rotational computed laminography (CL) has broad application potential in three-dimensional imaging of plate-like objects, as it only needs x-ray to pass through the tested object in the thickness direction during the imaging process. In this study, a square cross-section FOV rotational CL (SC-CL) was proposed. Then, the FDK-type analytical reconstruction algorithm applicable to the SC-CL was derived. On this basis, the proposed method was validated through numerical experiments.

## 1. INTRODUCTION

Computed tomography (CT) technology is widely used in the industrial field as a non-destructive testing technique [1, 2]. However, when imaging plate-like objects such as fossils, paintings, composite panels, printed circuit boards (PCB), the commonly used circular cone-beam CT is hard to obtain high-precision 3D images due to the limitation of the imaging space and radiation source energy [3, 4]. At the same time, computed laminography (CL) technology only requires rays to pass through the tested object in the thickness direction, so it has great potential for application in obtaining internal features of plate-like objects [5]. In the early days, CL was mainly through the imaging of object focal plane to obtain the internal information of the object. With the development of computer and digital detector technology and the improvement of CL reconstruction algorithms, CL was able to obtain three-dimensional images of objects like CT [6, 7].

According to the difference of scan trajectory, CL can be divided into translation CL, rotational CL, and swing CL, among which rotational CL is widely used due to its strong adaptability and rich projection information [8]. rotational CL is analogous to circular cone-beam CT in terms of the scanning geometry: the detector and x-ray source rotate 360° around the rotation axis to collect projection information. But the angles between the beam's principal ray and the rotation axis (i.e., tilt angle $\alpha$ in Fig. 1) in both systems are different. In circular cone-beam CT, the beam's principal ray is perpendicular to the rotation axis ($\alpha$=90°), while the angle in rotational CL is less than 90° ($\alpha$<90°), as shown in Fig. 1. And it is this characteristic enables x-ray to only pass the plate-like object in the thickness direction during the 360 ° scanning process [9].

In this study, we proposed a square cross-section FOV rotational CL and its analytical reconstruction algorithm was derived and verified through numerical experiments. Finally, the proposed method is applied to defect detection of actual printed circuit boards.

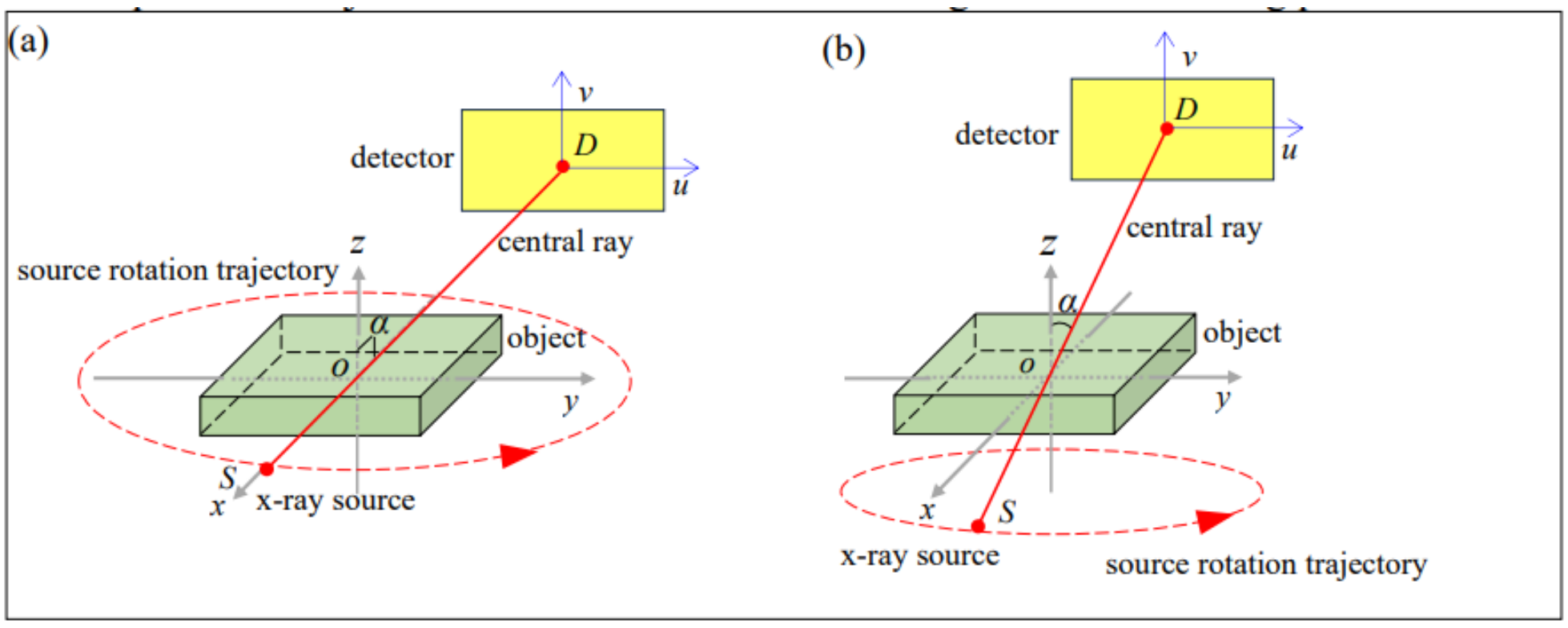

**Fig. 1** Imaging geometric diagram of CT and CL: (a) circular cone-beam CT, (b) rotational CL. *O* is the center of the object, *S* is the source, *D* is the center of the detector, *v* and *u* are the vertical and horizontal axes of the detector local coordinate system, respectively. The ray passing through *S*, *O*, and *D* in sequence is called central ray.

## 2. METHODOLOGY

Figure 2 shows system configuration of SC-CL, its detector is horizontally arranged and its orientation keeps constant during rotation. In this way, the cross-section of its field of view (FOV) is square. Because the imaging geometry SC-CL is different from CBCT, the classical FDK algorithm cannot be directly used. Therefore, an analytical reconstruction method specifically for SC-CL is necessary for efficient and good reconstruction.

As shown in Fig. 2 (a), where a global coordinate system *O-xyz* is defined with *z* being the rotation axis and the origin *O* is the intersection of axis *z* and the line connecting the source (*S*) and the center of detector (*D*). The Zenith angle $\alpha$ is referred as the CL tilt angle. Plane *E* is the plane where the detector locates and *O*' is the intersection of plane *E* and rotation axis. *S*' is the projection of *S* on plane *E*. Fig. 2 (b) is the 2D schematic diagram from top view on the plane *E*. There is a native coordinate system *D-uv* in the detector. During rotation, the directions of axis *u* and *v* in *D-uv* remain parallel to the axes *x* and *y* respectively.

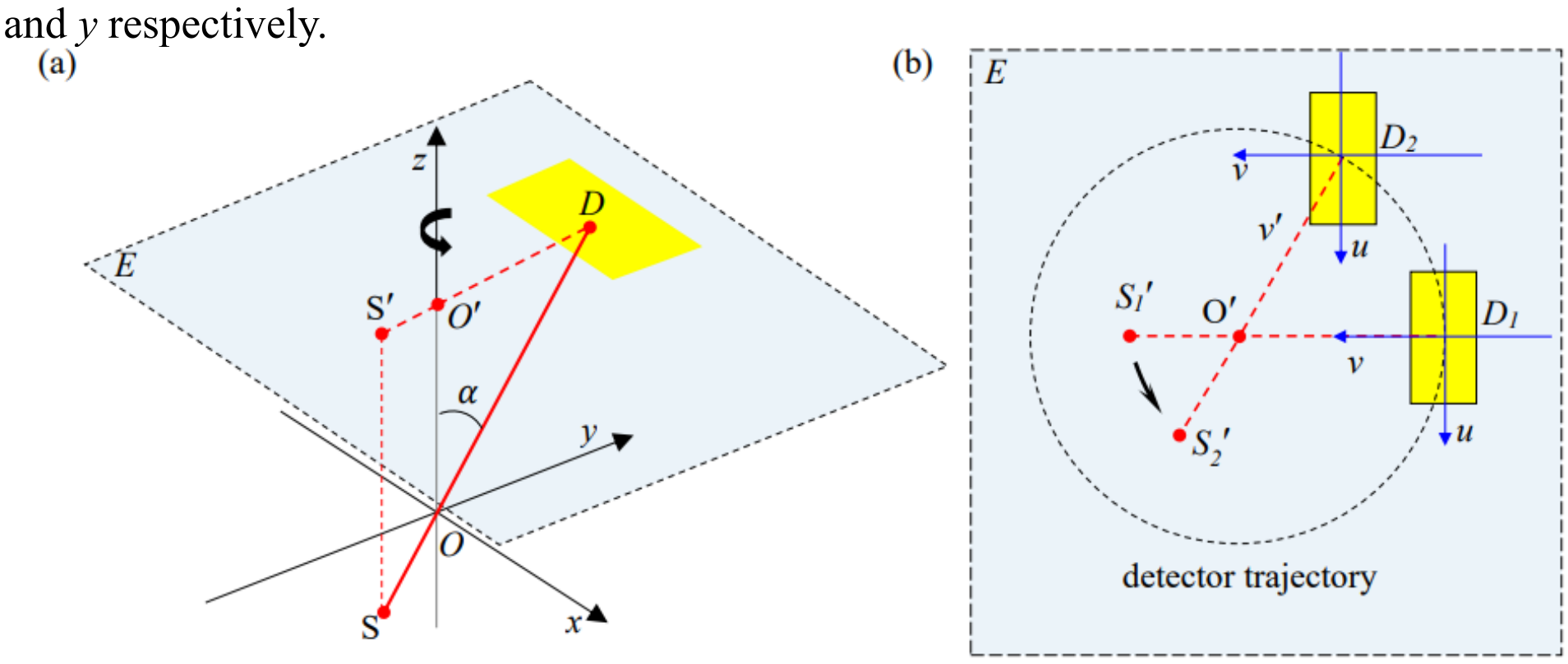

**Fig. 2** Imaging diagram of SC-CL: (a) 3D schematic diagram, (b) schematic diagram on plane *E*, $S_1'/D_1$ and $S_2'/D_2$ are locations of source/detector at two different projection angles, respectively.

### （A）Formulation of analytical reconstruction on a virtual 2D problem

To derive the reconstruction formula for SC-CL, we follow the idea of FDK method and start from a 2D filtered back projection (FBP) reconstruction on plane *E*. Firstly, a rotational coordinate system *D-u*'*v*' is configured on the detector with axis *v*' points to the rotation axis during rotation to form a 2D virtual problem. For convenience, we define the angle between axis *v*' and *v* to be our projection angle $\beta$. The relation between *D-uv* and *D-u*'*v*' at projection angle $\beta$ is:

$$\begin{cases} u' = u\cos\beta - v\sin\beta \\ v' = u\sin\beta + v\cos\beta \end{cases} \tag{1}$$

As shown in Fig. 3, on the detector plane *E*, if we regard point *S*' as the x-ray source in the 2D problem, projection data along *v*' at a certain $v' = v'^{*}$ gives a standard view of a fan-beam CT. Hence, we can apply a FBP reconstruction algorithm in this situation:

$$f(x,y)=\frac{1}{2}\int_0^{2\pi}(\frac{|S'D'|}{|S'R'|})^2\int_{u'_{\min}}^{u'_{\max}}\frac{|S'D'|}{\sqrt{|S'D'|^2+u'^2}}p_\beta(u',v'^*)h(u'-u'^*)\mathrm{d}u'\mathrm{d}\beta=$$
$$\frac{1}{2}\int_0^{2\pi}(\frac{|S'D|-v'^*}{|S'R'|})^2\int_{u'_{\min}}^{u'_{\max}}\frac{|S'D|-v'^*}{\sqrt{(|S'D|-v'^*)^2+u'^2}}p_\beta(u',v'^*)h(u'-u'^*)\mathrm{d}u'\mathrm{d}\beta \tag{2}$$

where $(x,y)$ is the coordinates of a reconstruction point $R$. $D'$ is the point on $v'$ axis with $v = v'^*$. $|S'R'|$ is the projected distance of $|S'R|$ on $|S'D'|$. $u'^*$ and $v'^*$ are the projection positions of point $R$ on detector. $|S'D|$ and $|S'O'|$ are the distance from source $S'$ to the detector center $D$ and origin $O'$, respectively. $p_\beta(u',v'^*)$ represents projection data at location $(u',v'^*)$, and $h(u)$ is a ramp filter.

In Eq. (2), the FBP filtering is performed along $u'$ axis. However, the projection data in SC-CL is recorded along $u$ and $v$ axes. Therefore, a reconstruction formula for SC-CL will be derived based on Eq. (2).

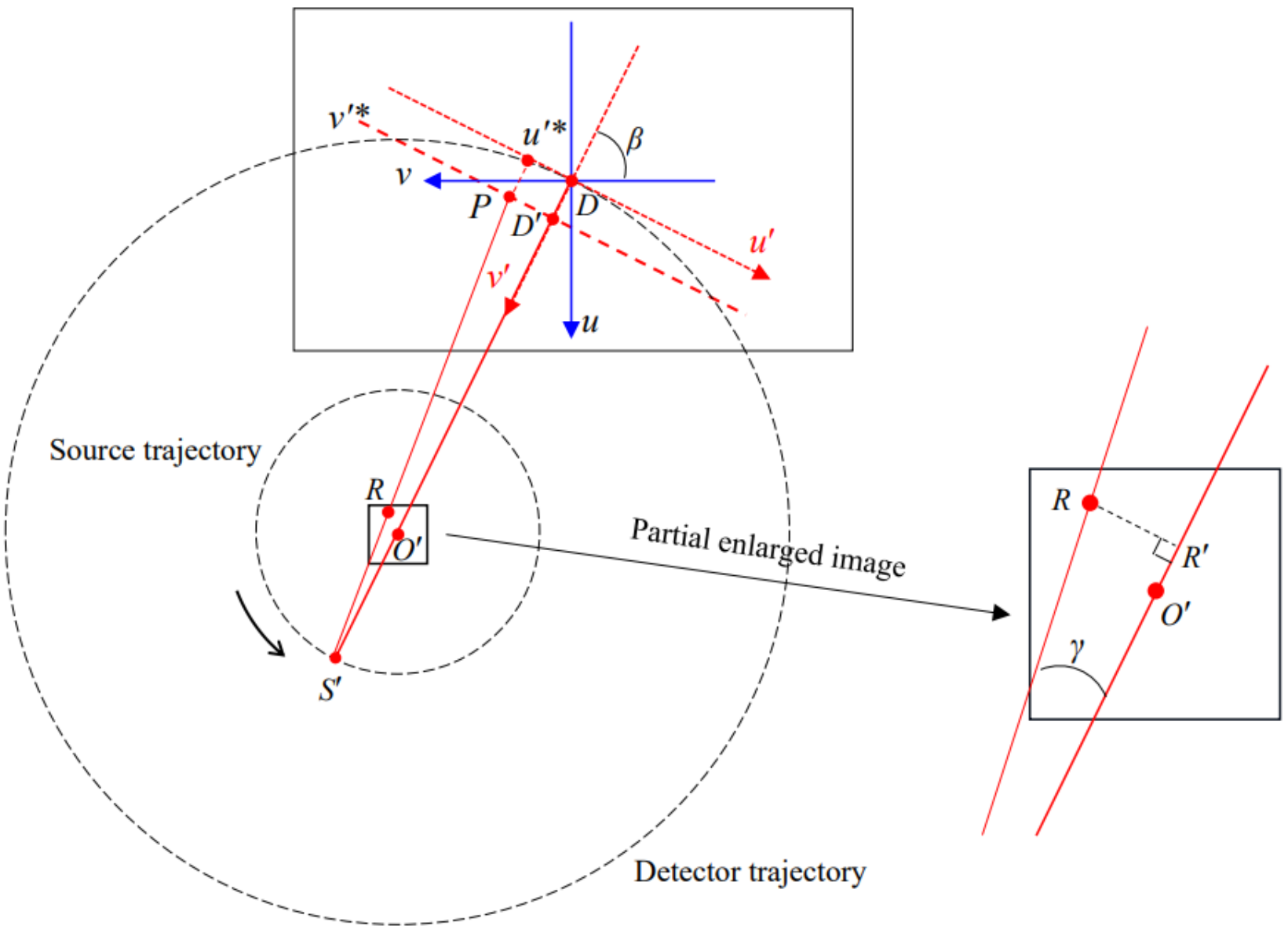

**Fig. 3** Geometry in 2D situation with $D$, $O'$, and $S'$ representing the locations of detector center, isocenter and source respectively, and $P$ is the projection of $R$ on detector.

Firstly, to simplify the derivation, let's define:

$$p^{\text{filtered}}_{(\beta,u'^*,v'^*)}(u')\triangleq\int_{u'_{\min}}^{u'_{\max}}\frac{|S'D|-v'^*}{\sqrt{(|S'D|-v'^*)^2+u'^2}}p_\beta(u',v'^*)h(u'^*-u')\mathrm{d}u' \tag{3}$$

It can be further expressed as:

$$p^{\text{filtered}}_{(\beta,u'^*,v'^*)}(u',v')=\int_{v'_{\min}}^{v'_{\max}}\delta(v'-v'^*)\mathrm{d}v'\int_{u'_{\min}}^{u'_{\max}}\frac{|S'D|-v'^*}{\sqrt{(|S'D|-v'^*)^2+u'^2}}p_\beta(u',v')h(u'^*-u')\mathrm{d}u' \tag{4}$$

Substituting Eq. (1) into Eq. (4) gives:

$$p^{\text{filtered}}_{(\beta,u^*,v^*)}(u,v)=\int_{v_{\min}}^{v_{\max}}\delta(u\sin\beta+v\cos\beta-u^*\sin\beta-v^*\cos\beta)\mathrm{d}v$$
$$\int_{u_{\min}}^{u_{\max}}\frac{|S'D|-u^*\sin\beta-v^*\cos\beta}{\sqrt{(|S'D|-u^*\sin\beta-v^*\cos\beta)^2+(u\cos\beta-v\sin\beta)^2}}g_\beta(u,v) \tag{5}$$
$$h(u^*\cos\beta-v^*\sin\beta-u\cos\beta+v\sin\beta)\mathrm{d}u$$

where $g_\beta(u,v)$ is the projection data on physical detector grids.

In Eq. (5), according to the scaling property of Dirac delta function, we obtain

$$\delta(u\sin\beta+v\cos\beta-u^*\sin\beta-v^*\cos\beta)=\frac{1}{|\cos\beta|}\delta(\frac{u\sin\beta-u^*\sin\beta-v^*\cos\beta}{\cos\beta}+v) \tag{6}$$

By substituting Eq. (6) into Eq. (5), we obtain:

$$\begin{aligned} p^{\text{filtered}}_{(\beta,u^*,v^*)}(u,v) &= \frac{1}{|\cos\beta|}\int_{v_{\min}}^{v_{\max}}\delta(\frac{u\sin\beta-u^*\sin\beta-v^*\cos\beta}{\cos\beta}+v)\mathrm{d}v \\ &\int_{u_{\min}}^{u_{\max}}\frac{|S'D|-u^*\sin\beta-v^*\cos\beta}{\sqrt{(|S'D|-u^*\sin\beta-v^*\cos\beta)^2+(u\cos\beta-v\sin\beta)^2}}g_\beta(u,v)h(u^*\cos\beta-v^*\sin\beta-u\cos\beta+v\sin\beta)\mathrm{d}u \end{aligned} \tag{7}$$

From $v+\frac{u\sin\beta-u^*\sin\beta-v^*\cos\beta}{\cos\beta}=0$, we obtain

$$v=\frac{u^*\sin\beta+v^*\cos\beta-u\sin\beta}{\cos\beta} \tag{8}$$

By substituting Eq. (8) into Eq. (7), we obtain:

$$\begin{aligned} p^{\text{filtered}}_{(\beta,u^*,v^*)}(u) &= \frac{1}{|\cos\beta|}\int_{u_{\min}}^{u_{\max}}\frac{|S'D|-u^*\sin\beta-v^*\cos\beta}{\sqrt{(|S'D|-u^*\sin\beta-v^*\cos\beta)^2+(\frac{u-u^*\sin^2\beta-v^*\cos\beta\sin\beta}{\cos\beta})^2}} \\ &g_\beta(u,\frac{u^*\sin\beta+v^*\cos\beta-u\sin\beta}{\cos\beta})h(\frac{u^*-u}{\cos\beta})\mathrm{d}u \end{aligned} \tag{9}$$

According to Fourier transforming property: $F\{f(at)\}=\frac{1}{|a|}F(\frac{w}{a})$, we obtain

$$h(\frac{u^*-u}{\cos\beta})=(\cos\beta)^2h(u^*-u) \tag{10}$$

By substituting Eq (10) into Eq. (9), we obtain:

$$\begin{aligned} p^{\text{filtered}}_{(\beta,u^*,v^*)}(u) &= |\cos\beta|\int_{u_{\min}}^{u_{\max}}\frac{|S'D|-u^*\sin\beta-v^*\cos\beta}{\sqrt{(|S'D|-u^*\sin\beta-v^*\cos\beta)^2+(\frac{u-u^*\sin^2\beta-v^*\cos\beta\sin\beta}{\cos\beta})^2}} \\ &g_\beta(u,\frac{u^*\sin\beta+v^*\cos\beta-u\sin\beta}{\cos\beta})h(u^*-u)\mathrm{d}u \end{aligned} \tag{11}$$

By combining Eq. (11) and (2), we obtain an FBP-type reconstruction formula for SC-CL:

$$\begin{aligned} f(x,y) &= \frac{1}{2}\int_0^{2\pi}(\frac{|S'D|-v'^*}{|S'R'|})^2 d\beta\int_{u_{\min}}^{u_{\max}}\frac{|S'D|-u^*\sin\beta-v^*\cos\beta}{\sqrt{(|S'D|-u^*\sin\beta-v^*\cos\beta)^2+(\frac{u-u^*\sin^2\beta-v^*\cos\beta\sin\beta}{\cos\beta})^2}} \\ &g_\beta(u,\frac{u^*\sin\beta+v^*\cos\beta-u\sin\beta}{\cos\beta})|\cos\beta|h(u^*-u)\mathrm{d}u \end{aligned} \tag{12}$$

### 3.2 Extension to a 3D scenario

In 3D situation, the *z* dimension must be considered during the reconstruction. Same as the derivation of FDK algorithm for CBCT, when extending FBP algorithm from 2D to 3D, two items in FBP need to be recalculated.

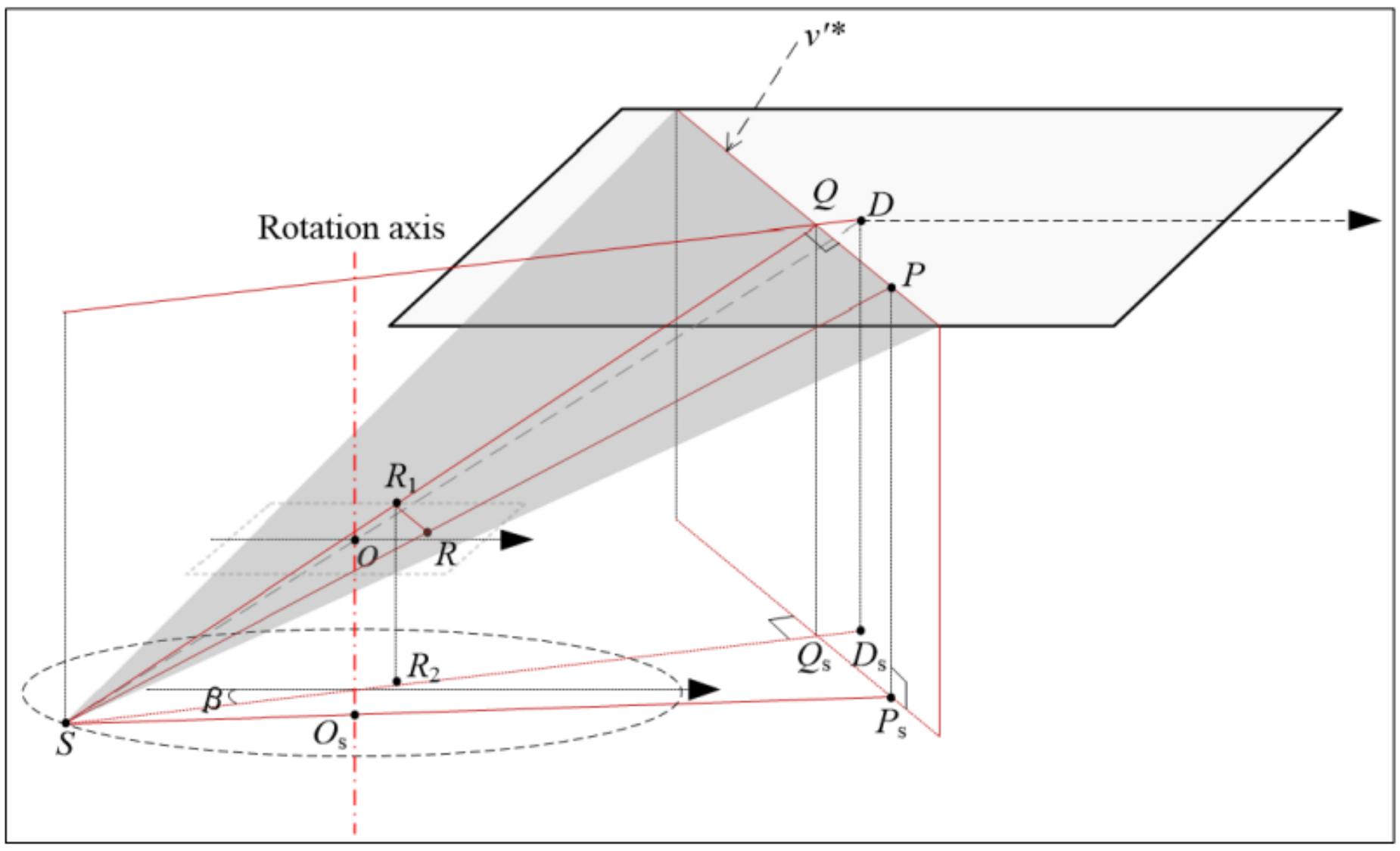

**Fig. 4** Geometry in 3D situation, $R$ is the reconstruction point and its projection on detector is $P$. $Q$ is the point on $v'$ axis with $v' = v'^*$ .

The first one is the weighting factor prior to the filtering operation (recorded as $\eta_1$). According to Eq. (12), the expression of $\eta_1$ in FBP is

$$\eta_1 = \frac{|S'D'|}{\sqrt{|S'D'|^2 + u'^2}} = \frac{|S'D| - u^* \sin\beta - v^* \cos\beta}{\sqrt{(|S'D| - u^* \sin\beta - v^* \cos\beta)^2 + (\frac{u - u^* \sin^2\beta - v^* \cos\beta \sin\beta}{\cos\beta})^2}} \tag{13}$$

Physically speaking, in 2D case, $\eta_1$ stands for the cosine of fan angle of reconstruction point $R$. In 3D case, as shown in Fig. 4, fan angle of reconstruction point $R$ is $\angle P_s SD_s$ . Meanwhile, the influence of cone angle (i.e., $\angle PSP_s$ in Fig. 4) needs to be considered. Therefore, the expression of $\eta_1$ is

$$\eta_1 = \cos\angle D_s SP_s * \cos\angle PSP_s = \cos\angle O_s SP_s * \cos\angle PSP \tag{14}$$

According to the cosine theorem, Eq. (14) can be written as:

$$\eta_1 = \cos\angle O_s SP_s * \cos\angle PSP_s = \frac{|SP_s|^2 + |SO_s|^2 - |O_s P_s|^2}{2 \cdot |SP_s| \cdot |SO_s|} \frac{|SP_s|}{|SP|} = \frac{|SP_s|^2 + |SO_s|^2 - |O_s P_s|^2}{2 \cdot |SP| \cdot |SO_s|} \tag{15}$$

In global coordinate *O-xyz*, the coordinate of point $O_s$, $S$, $P$, $P_s$ are $(0, 0, -|SO|\cos\alpha)$ $(|SO|\sin\alpha\sin\beta, -|SO|\sin\alpha\cos\beta, -|SO|\cos\alpha)$ , $(-|OD|\sin\alpha\sin\beta + u, |OD|\sin\alpha\cos\beta - v, |OD|\cos\alpha)$ , $(-|OD|\sin\alpha\sin\beta + u, |OD|\sin\alpha\cos\beta - v, -|OD|\cos\alpha)$ , Eq.(15) can be expressed as:

$$\eta_1 = \frac{|SD|\sin\alpha - v\cos\beta - u\sin\beta}{\sqrt{|SD|^2 - 2|SD|\sin\alpha(u\sin\beta + v\cos\beta) + u^2 + v^2}} \tag{16}$$

By substituting Eq. (8) into Eq. (16), we obtain:

$$\eta_1 = \frac{|SD|\sin\alpha - u^*\sin\beta - v^*\cos\beta}{\sqrt{|SD|^2 - 2|SD|\sin\alpha(u^*\sin\beta + v^*\cos\beta) + u^2 + (\frac{u^*\sin\beta + v^*\cos\beta - u\sin\beta}{\cos\beta})^2}} \tag{17}$$

The second one is the weighting factor for backprojection (recorded as $\eta_2$ ), According to Eq. (12), the expression of $\eta_2$ in FBP is

$$\eta_2 = (\frac{|S'D| - v'^*}{|S'R'|})^2 \tag{18}$$

Physically speaking, $\eta_2$ is determined by the source-to-detector distance $|S'D| - v'^*$ and the projected distance $|S'R'|$ between the source and the reconstruction point on the central ray. Therefore, as shown in Fig.4, its expression in 3D case is

$$\eta_2 = (\frac{|SQ|}{|SR_1|})^2 \tag{19}$$

According to triangle similarity theorem, we can obtain:

$$\frac{|SQ|}{|SR_1|} = \frac{|QQ_s|}{|R_1R_2|} = \frac{|SD| \cdot \cos(\alpha)}{z + |SO| \cdot \cos(\alpha)} \tag{20}$$

where $z$ is the $z$ coordinate of reconstruction point

Therefore, the Eq. (19) can be written as：

$$\eta_2 = (\frac{|SQ|}{|SR_1|})^2 = (\frac{|SD| \cdot \cos(\alpha)}{z + |SO| \cdot \cos(\alpha)})^2 \tag{21}$$

Replace the $\eta_1$ and $\eta_2$ by Eqs. (17) and (21), the FDK-type reconstruction formula in SC-CL can be obtained.

$$\begin{aligned} f(x,y,y) &= \frac{1}{2}\int_0^{2\pi} (\frac{|SD| \cdot \cos(\alpha)}{z + |SO| \cdot \cos(\alpha)})^2 d\beta \\ &\int_{u_{\min}}^{u_{\max}} \frac{|SD|\sin\alpha - u^*\sin\beta - v^*\cos\beta}{\sqrt{|SD|^2 - 2|SD|\sin\alpha(u^*\sin\beta + v^*\cos\beta) + u^2 + (\frac{u^*\sin\beta + v^*\cos\beta - u\sin\beta}{\cos\beta})^2}} \\ &g_\beta(u, \frac{u^*\sin\beta + v^*\cos\beta - u\sin\beta}{\cos\beta})|\cos\beta| h(u^* - u) \mathrm{d}u \end{aligned} \tag{22}$$

The implementation steps for the proposed algorithm can be summarized into three steps:

**1) Preweighting:** Multiply the two-dimensional projection data by a weighting factor for correction. The expression of the weight is: $|\cos\beta| \frac{|SD|\sin\alpha - u^*\sin\beta - v^*\cos\beta}{\sqrt{|SD|^2 - 2|SD|\sin\alpha(u^*\sin\beta + v^*\cos\beta) + u^2 + (\frac{u^*\sin\beta + v^*\cos\beta - u\sin\beta}{\cos\beta})^2}}$.

**2) Filtration:** In numerical implementation, to lower discretization error and avoid $\cos\beta = 0$ at $\beta = \frac{\pi}{2}$ or $\frac{3\pi}{2}$, , We divide $[0, 2\pi)$ into four parts, $[-\frac{\pi}{4}, \frac{\pi}{4})$, $[\frac{\pi}{4}, \frac{3\pi}{4})$, $[\frac{3\pi}{4}, \frac{5\pi}{4})$ and $[\frac{5\pi}{4}, \frac{7\pi}{4})$.

Specially, when $\beta \subseteq [-\frac{\pi}{4}, \frac{\pi}{4}) \cup [\frac{3\pi}{4}, \frac{5\pi}{4})$, $v = \frac{1}{\cos\beta}(u^*\sin\beta + v^*\cos\beta - u\sin\beta)$ is adopted, which means we do filtration along the $u$-axis. When $\beta \subseteq [\frac{\pi}{4}, \frac{3\pi}{4}) \cup [\frac{5\pi}{4}, \frac{7\pi}{4})$, $u = \frac{1}{\sin\beta}(u^*\sin\beta + v^*\cos\beta - v\cos\beta)$ , which means we do filtration is along the $v$-axis. Though we managed to simplify the reconstruction step to a 1D filtration along the natural axis of $u$ or $v$, we still need some interpolation step since the projection data is not on grid. Fortunately, 1D interpolation is enough and it is easy to be done.

**3) weighted back projection:** The 3D back-projection weighted by $(\frac{|SD| \cdot \cos(\alpha)}{|SO| \cdot \cos(\alpha) + z})^2$ is similar to other FDK type reconstructions.

## 3. EXPERIMENTS AND RESULTS

To verify the proposed method (referred to below as CL-FDK), we simulated a SC-CL system with horizontal and fixed-orientation detector. In the simulation, the ray- and voxel-driven models were chosen as forward- and back-projectors, respectively. A PCB phantom as shown in Fig. 5 is used. The phantom contains three copper circuit layers that are interconnected. The detailed imaging parameters were: the title angle $\beta$ was 45°, the distances from source to origin and detector center were 45.790mm and 194.580mm, respectively; The detector was simulated with a 768×768 array and a 0.17×0.17 $mm^2$ pixel size, and 256 projection images were acquired. The reconstruction image grids are 300×300×80 with a 0.07×0.07×0.07 $mm^3$ voxel size.

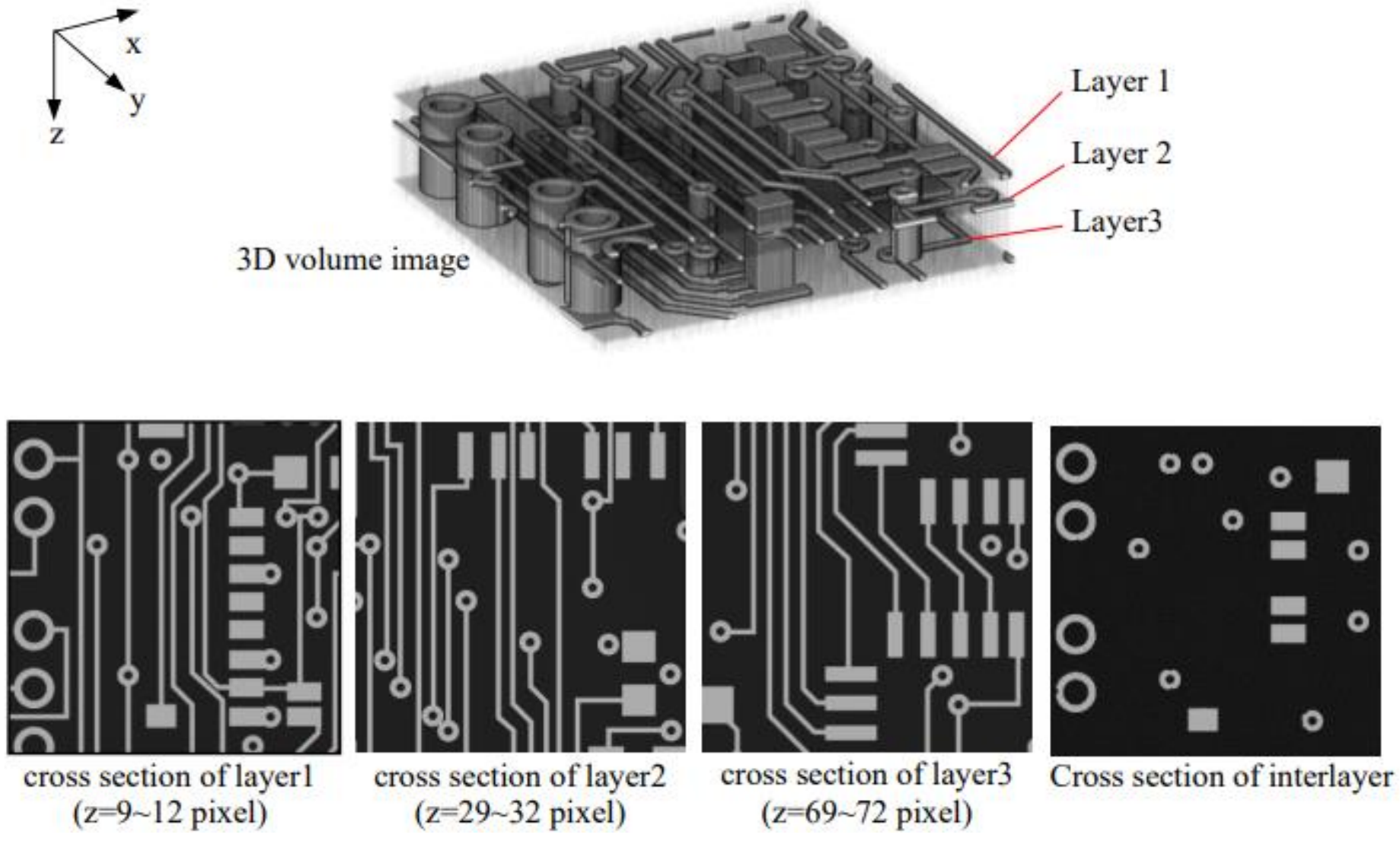

**Fig. 5** Numerical PCB phantom for simulation study.

To quantitatively evaluate the quality of the reconstrued PCB images, two metrics were used to measure the similarity between the reconstructed image and the reference image: root mean square error (RMSE) and mean structural similarity index (MSSIM). The smaller the RMSE, the better the reconstruction quality; The larger the MSSIM, the better the reconstruction quality.

For comparison, the reconstruction results by the PT-FDK method and the simultaneous iterative reconstruction technique (SIRT) were also presented, with the SIRT iteration count set to 200. The reconstructed results of three algorithms are shown in Fig. 6. It can be seen that all three methods can reconstruct the main structural features in the phantom. It can also be noted that the error in CL-FDK result is smaller than PT-FDK.

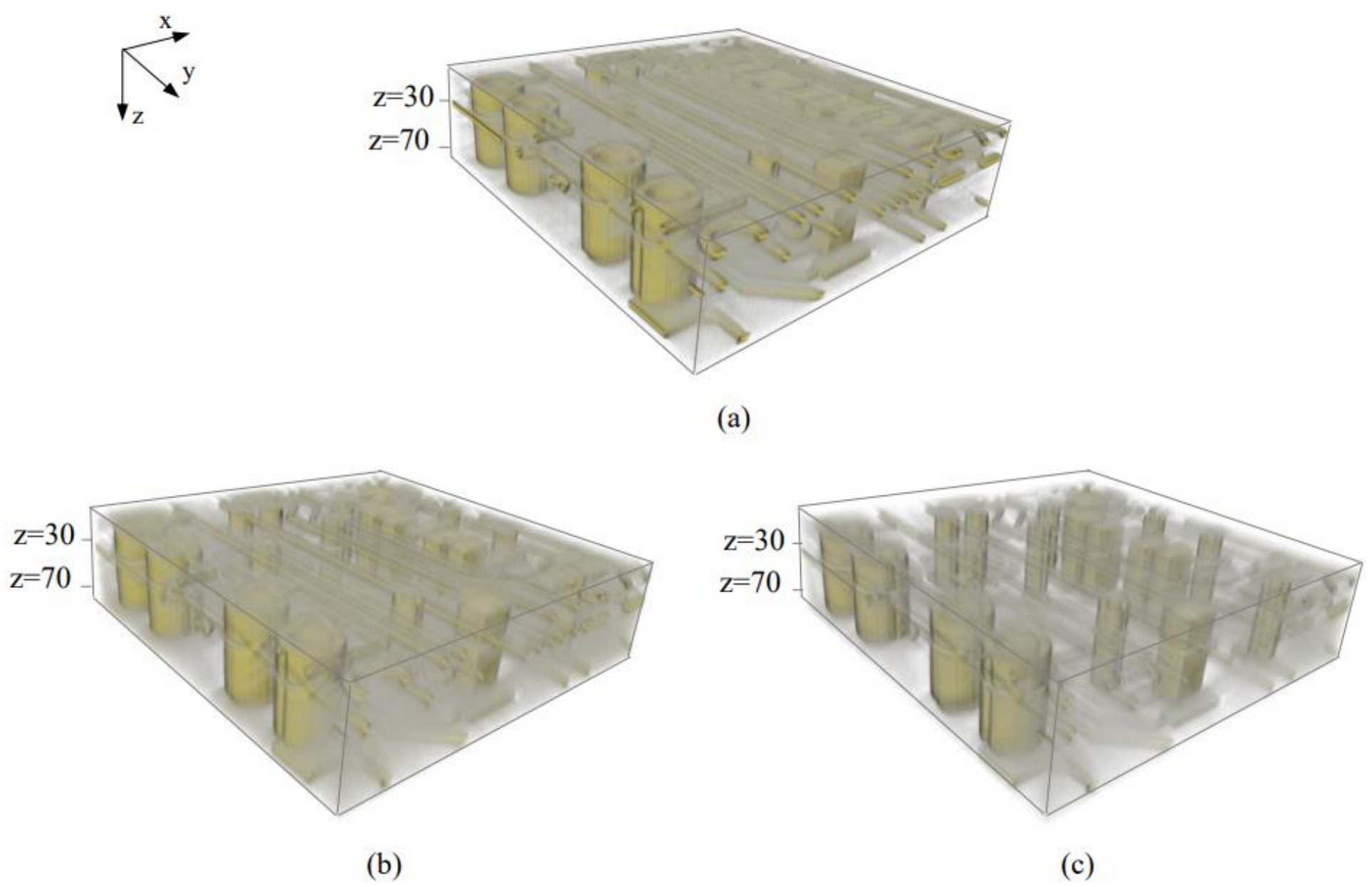

**Fig. 6** Volumetric rendering of reconstruction results from three algorithms: (a) SIRT, (b) PT-FDK, (c) CL-FDK.

Fig.7 shows the RMSE and SSIM results for different reconstruction methods. It can be seen that the SIRT reconstruction method is the best, followed by CL-FDK, and the PT-FDK is the worst. As a filtering backprojection algorithm, the PT-FDK algorithm has lower accuracy than CL-FDK mainly because PT-FDK requires out of plane interpolation of the projection image during the transformation process, and the additional interpolation operation not only increases the computational load but also brings interpolation errors.

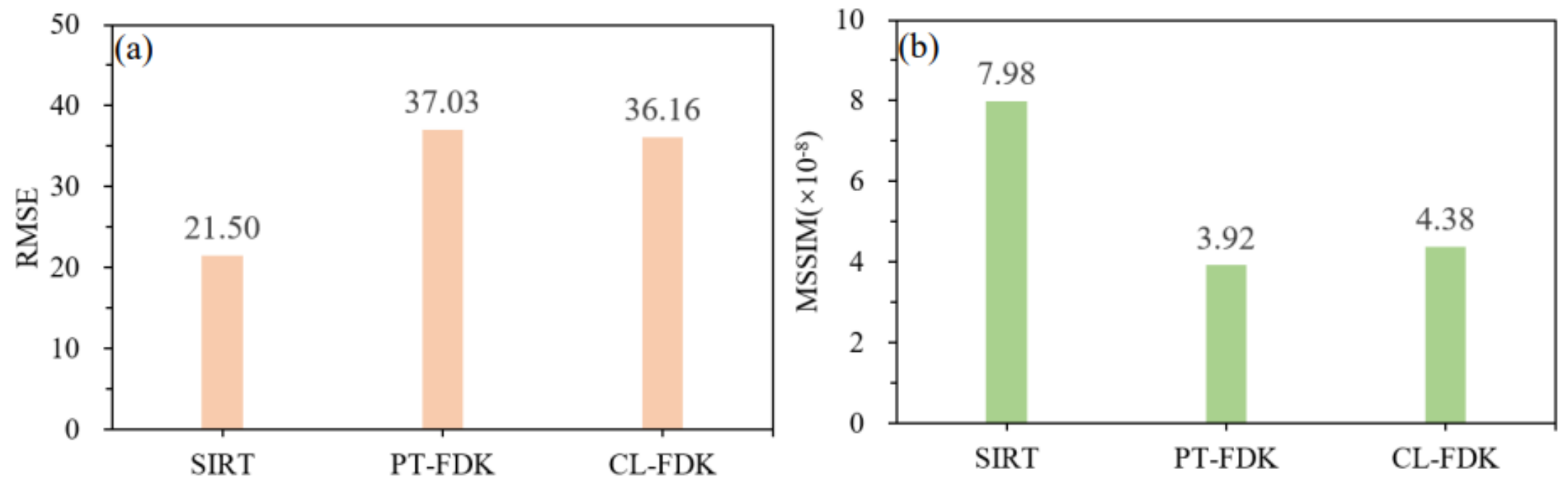

**Fig. 7** Comparison of evaluation indicators: (a) RMSE, (b) MSSIM.

## 4. DISCUSSION AND CONCLUSION

This paper proposed a SC-CL with horizontal and fixed-orientation detector, and then derived its analytical reconstruction algorithm. The algorithm is validated and superior performance over PT-FDK is obtained.

It should be pointed out that both the typical SIRT iterative reconstruction algorithms and analytical reconstruction algorithms suffer from interlayer aliasing artifacts, which have not yet been fully resolved in the field. We are to conduct further research on this issue in the future. Especially, deep learning methods have made significant breakthroughs in the field of x-ray CT, combining CL-FDK with deep learning methods to reduce interlayer aliasing artifacts in SC-CL images might be a promising approach.

## References

[1] Wang T, Nakamoto K, Zhang H, et al. Reweighted Anisotropic Total Variation Minimization for Limited-Angle CT Reconstruction[J]. IEEE transactions on nuclear science, 2017,64(10):2742-2760.

[2] Zheng A, Gao H, Zhang L, et al. A dual-domain deep learning-based reconstruction method for fully 3D sparse data helical CT[J]. Phys Med Biol, 2020,65(24):245030.

[3] Shi L, Wei C, Jia T, et al. An automatic measurement method of PCB stub based on rotational computed laminography imaging[J]. IEEE transactions on nuclear science, 2023,70(9):1.

[4] Ghandourah E E, Hamidi S H A, Mohd Salleh K A, et al. Evaluation of Welding Imperfections with X-ray Computed Laminography for NDT Inspection of Carbon Steel Plates[J]. Journal of nondestructive evaluation, 2023,42(3).

[5] Fisher S L, Holmes D J, Jørgensen J S, et al. Laminography in the lab: imaging planar objects using a conventional x-ray CT scanner[J]. Measurement science & technology, 2019,30(3):35401.

[6] Zhang Y, Yang M, Wu Y, et al. A New CL Reconstruction Method Under the Displaced Sample Stage Scanning Mode[J]. IEEE transactions on nuclear science, 2021,68(11):2574-2586.

[7] Li Y, Luo M, Fei X, et al. Online learning for DBC segmentation of new IGBT samples based on computed laminography imaging[J]. Discover Applied Sciences, 2024,6(4):145.

[8] Sun L, Zhou G, Qin Z, et al. A reconstruction method for cone-beam computed laminography based on projection transformation[J]. Measurement science & technology, 2021,32(4):45403.

[9] Zhao X, Jiang W, Zhang X, et al. Image reconstruction based on nonlinear diffusion model for limited-angle computed tomography[J]. Inverse problems, 2024.

[10] Ma G, Zhao X, Zhu Y, et al. Projection-to-image transform frame: a lightweight block reconstruction network for computed tomography[J]. Physics in medicine & biology, 2022,67(3):35010.

[11] Guo X, Zhang L, Xing Y. Study on analytical noise propagation in convolutional neural network methods used in computed tomography imaging[J]. Nuclear Science and Techniques, 2022,33(6).

[12] Ji P, Jiang Y, Zhao R, et al. Fusional laminography: A strategy for exact reconstruction on CL and CT information complementation[J]. NDT & E international : independent nondestructive testing and evaluation, 2024,141:102991.

[13] Lu J, Liu Y, Chen Y, et al. Cone beam computed laminography based on adaptive-weighted dynamic-adjusted relative total variation[J]. Nuclear instruments & methods in physics research. Section A, Accelerators, spectrometers, detectors and associated equipment, 2023,1051:168200.

[14] Sun Y, Han Y, Tan S, et al. Geometric parameters sensitivity evaluation based on projection trajectories for X-ray cone-beam computed laminography[J]. J Xray Sci Technol, 2023,31(2):423-434.

[15] Xing Y, Zhang L. A free-geometry cone beam CT and its FDK-type reconstruction[J]. Journal of x-ray science and technology, 2007,15(3):157-167.

[16] Yang H, Liang K, Kang K, et al. Slice-wise reconstruction for low-dose cone-beam CT using a deep residual convolutional neural network[J]. Nuclear science and techniques, 2019,30(4):53-61.

[17] Yang M, Wang G, Liu Y. New reconstruction method for x-ray testing of multilayer printed circuit board[J]. Optical Engineering, 2010,49(5):56501.

[18] Feldkamp L A, Davis L C, JKress J M. Practical cone-beam algorithm[J]. Journal of the Optical Society of America A, 1984,1(6):612-619.

[19] van Aarle W, Palenstijn W J, Cant J, et al. Fast and flexible X-ray tomography using the ASTRA toolbox[J]. Optics express, 2016,24(22):25129-25147.